# An algebra of automata which includes both classical and quantum entities

L. de Francesco Albasini    N. Sabadini    R.F.C. Walters

June 24, 2018

**Abstract**

We describe an algebra for composing automata which includes both classical and quantum entities and their communications. We illustrate by describing in detail a quantum protocol.

## 1 Introduction

The idea of this paper is to introduce quantum components into the algebra of automata introduced in [9]. This permits a compositional description of quantum protocols, in which quantum components interact with classical finite space components. The inclusion of finite state classical control adds conceptual clarity and precision to quantum protocols. Further, the undoubted subtlety of the interaction between the classical and quantum world justifies explicit description of the entities involved.

A mixed algebra of quantum and classical phenomena has already been introduced by Coecke and Pavlovic in [4], with further work in [5], following the categorical twist on quantum logic introduced in [1]. The idea of those works is to describe data flow in quantum protocols involving also classical measurements as expressions in a symmetric monoidal category with extra structure. Such a formulation yields geometric pictures (following [17],[12]) of the flow in protocols, as well as pictorial equations which may be used to prove correctness.

The current work introduces an extra level of description, also with an associated geometry, the geometry of the *classical and quantum entities* and *communications* between entities involved. We will indicate how the relation between the two levels (and pictures) is strongly analogous to the relation in concurrency theory between algebras of process and Petri nets [15].

At the level of entities the importance of the *distributive law* of tensor product over direct sum in making classical choices becomes evident. The situation is entirely analogous to classical Turing machines where an infinite state tape interacts with, and is controlled by, a finite state automaton (see the (non-compositional) description of Turing machines in [20]; and also [18],[19] for relations with the Blum-Shub-Smale theory of computable functions).

Another point of interest is that we find a common algebra for model checking (in which non-determinism and the state explosion are considered the main problem) and quantum computing (in which linearity and the expanded state space are the cited advantage).

Our automata are not to be confused with the quantum automata of [14] or [16], and hence we use the name ℂ-*automaton* for the general notion and *quantum or classical C-automaton* for those which represent respectively quantum or classical components.

We define a ℂ-automaton **Q** with a given set $A$ of "signals on the left interface", and set $B$ of "signals on the right interface" to consist of a finite dimensional complex vector space $V$ and a family of linear transformations $\varphi_{a,b} : V \to V$ ($a \in A, b \in B$). A quantum ℂ-automaton is one in which the space $V$ has the extra structure of an hermitian inner product, and in which the linear transformations are unitary transformations or orthogonal projections. A classical ℂ-automaton is one with the extra structure that the space $V$ is of the form $\mathbb{C}^X$ for a given finite set $X$ and for which the matrices of the linear transformations are zero-one matrices induced by binary relations on $X$.

The idea of [9] was to introduce two-sided automata, in order to permit operations analogous to the parallel, series and feedback of classical circuits, in particular in concurrency theory. We have more recently described a similar algebra for automata with *probability* in [7],[8].

As an illustration of the algebra we will give details of the teleportation protocol of [2].

## 2 ℂ-automata

**Definition 2.1** *Consider two finite alphabets $A$ and $B$. A ℂ-automaton **Q** with left interface $A$ and right interface $B$ consists of a finite dimensional complex vector space $V$ of states, and an $A \times B$ indexed family $\varphi = \varphi_{a,b(a \in A, b \in B)}$ of linear transformations from $V$ to $V$.*

**Definition 2.2** *A ℂ-automaton **Q** with the extra structure that the space $V$ is endowed with an hermitian inner product $< \mid >$ and for which the linear transformations are either unitary or orthogonal projections is called a* quantum ℂ-automaton.

**Definition 2.3** *A ℂ-automaton **Q** with the extra structure that the space $V$ is $\mathbb{C}^X$ for a given finite set $X$, and for which the linear transformations $\varphi_{a,b}$ are of the form $\varphi_{a,b}(e_x, e_y) = 0$ or $1$ ($x \in X$) (where $e_x$ ($x \in X$) is the standard basis of $\mathbb{C}^X$ defined by $e_x(y) = 1$ if $y = x$, and $0$ otherwise) is called a* classical (finite state) ℂ-automaton. *Note: we will often write just $x$ instead of $e_x$ for a basis element.*

The idea is that in a given state various transitions to other states are possible; the transitions that occur have effects, which we may think of a *signals*, on the two interfaces of the automaton, which signals are represented by letters in the alphabets. It is fundamental *not* to think of the letters in $A$ and $B$ in general as inputs or outputs, but rather signals induced by transitions of the automaton on the interfaces. For examples see a later section.

**Definition 2.4** *Consider a ℂ-automaton **Q** with interfaces $A$ and $B$. A behaviour of length $k$ of **Q** consists of a two words of length $k$, one $w_1 = a_1 a_2 \cdots a_k$ in $A^*$ and the other $w_2 = b_1 b_2 \cdots b_k$ in $B^*$ and a sequence of vectors*

$$\mathbf{x}_0, \mathbf{x}_1 = \varphi_{a_1,b_1}(\mathbf{x}_0), \mathbf{x}_2 = \varphi_{a_2,b_2}(\mathbf{x}_1), \cdots, \mathbf{x}_k = \varphi_{a_k,b_k}(\mathbf{x}_{k-1}).$$

# 3 Graphical representation

Although the definitions above are mathematically straightforward, in practice a graphical notation is more intuitive. Given a chosen basis for the state space of an automaton we may compress the description of an automaton with interfaces $A$ and $B$, which requires $A \times B$ matrices, into a single labelled graph, like the ones introduced in [9]. Further, expressions of automata in this algebra may be drawn as "tensor diagrams" also as in [9]. We indicate both of these matters by describing some examples.

## 3.1 Qubits

*Qubit automata* are a $\mathbb{C}$-automata with state space $\mathbb{C}^2$ which singly, or combined, form quantum automata. We will describe three particular qubit automata which will need for our discussion of teleportation. One of the qubit automata is a quantum automaton; the others will be combined to form a 2 qubit quantum automaton.

### 3.1.1 Qubit $Q_1$

Consider the alphabets $A_1 = \{\varepsilon, c, h, m_0, m_1\}$ and $B_1 = \{\varepsilon, \neg\}$. Then $\mathbf{Q}_1$ is the automaton with left interface $A_1$ and right interface $B_1$, state space $\mathbb{C}^2$ and transition matrices

$$\varphi_{\varepsilon,\varepsilon} = \begin{bmatrix} 1 & 0 \\ 0 & 1 \end{bmatrix}, \; \varphi_{c,\neg} = \begin{bmatrix} 0 & 0 \\ 0 & 1 \end{bmatrix}$$

$$\varphi_{c,\varepsilon} = \begin{bmatrix} 1 & 0 \\ 0 & 0 \end{bmatrix}, \; \varphi_{h,\varepsilon} = \frac{1}{\sqrt{2}} \begin{bmatrix} 1 & 1 \\ 1 & -1 \end{bmatrix}$$

$$\varphi_{m_0,\varepsilon} = \begin{bmatrix} 1 & 0 \\ 0 & 0 \end{bmatrix}, \; \varphi_{m_1,\varepsilon} = \begin{bmatrix} 0 & 0 \\ 0 & 1 \end{bmatrix}.$$

The other four transition matrices are zero matrices.

The intention behind these matrices is as follows: $\mathbf{Q}_1$ may do a transition labelled $\varepsilon, \varepsilon$ (*idle transition*); $\mathbf{Q}_1$ may receive a signal $h$ and perform a transition determined by the unitary Hadamard matrix; $\mathbf{Q}_1$ may receive a signal $c$ (*do* $\mathrm{C_{not}}$) and if it is in state 1 pass on a signal $\neg$ with the intention to perform a *not* on another qubit; the signal $m_0$ means that a *measurement* with result 0 has occurred on $\mathbf{Q}_1$; the signal $m_1$ means that a *measurement* with result 1 has occurred on $\mathbf{Q}_1$. All this information may be put in the following diagram, noting that (i) the basis elements of $\mathbb{C}^2$ are called 0 and 1, and occur in the diagram as vertices, (ii) labels of transitions indicate which matrix is imvolved, (iii) the absence of an edge from $i$ to $j$ means that the $i,j$th element of the matrix is 0, (iv) we have in any case omitted loops labelled $\varepsilon, \varepsilon$, (v) we have included the value of the matrix element only when it is not 1.

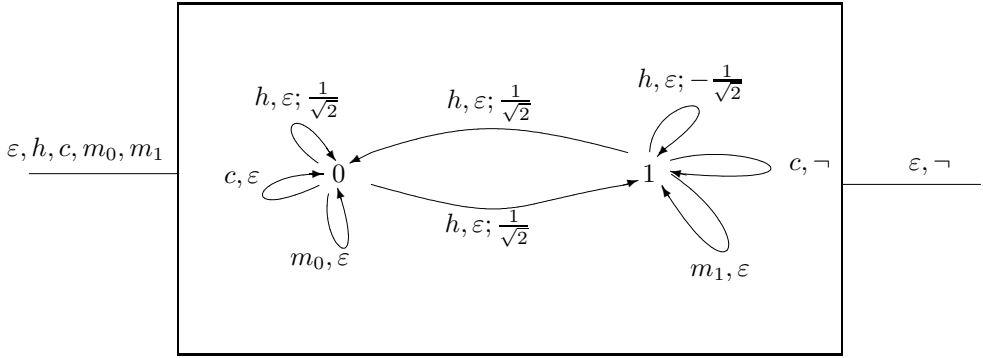

### 3.1.2 Qubit $Q_2$

Consider the alphabets $A_2 = \{\varepsilon, \neg\} \times \{\varepsilon, m_0, m_1\} = A_{21} \times A_{22} = B_1 \times A_{22}$ and $B_2 = \{\varepsilon\}$. Then $\mathbf{Q}_2$ is the automaton with left interface $A_2$ and right interface $B_2$, state space $\mathbb{C}^2$ and transition matrices

$$\varphi_{(\varepsilon,\varepsilon),,\varepsilon} = \begin{bmatrix} 1 & 0 \\ 0 & 1 \end{bmatrix}, \; \varphi_{(\neg,\varepsilon),\varepsilon} = \begin{bmatrix} 0 & 1 \\ 1 & 0 \end{bmatrix}$$

$$\varphi_{(\varepsilon,m_0),\varepsilon} = \begin{bmatrix} 1 & 0 \\ 0 & 0 \end{bmatrix}, \; \varphi_{(\varepsilon,m_1),\varepsilon} = \begin{bmatrix} 0 & 0 \\ 0 & 1 \end{bmatrix}.$$

The remaining matrices are zero.

The intention behind these matrices is as follows: $\mathbf{Q}_2$ may do a transition labelled $\varepsilon, \varepsilon$ (*idle transition*); $\mathbf{Q}_2$ may receive a signal $\neg$ and perform a *not* transition; the signal $m_0$ means that a *measurement* with result 0 has occurred on $\mathbf{Q}_2$; the signal $m_1$ means that a *measurement* with result 1 has occurred on $\mathbf{Q}_2$.

### 3.1.3 Qubit $Q_3$

Consider the alphabets $A_3 = \{\varepsilon\}$, and $B_3 = \{\varepsilon, 00, 01, 10, 11\}$. Then $\mathbf{Q}_3$ is the automaton with left interface $A_3$ and right interface $B_3$, state space $\mathbb{C}^2$ and transition matrices

$$\varphi_{\varepsilon,\varepsilon} = \begin{bmatrix} 1 & 0 \\ 0 & 1 \end{bmatrix},$$

$$\varphi_{\varepsilon,00} = \begin{bmatrix} 1 & 0 \\ 0 & 1 \end{bmatrix}, \; \varphi_{\varepsilon,10} = \begin{bmatrix} 1 & 0 \\ 0 & -1 \end{bmatrix},$$

$$\varphi_{\varepsilon,01} = \begin{bmatrix} 0 & 1 \\ 1 & 0 \end{bmatrix}, \; \varphi_{\varepsilon,11} = \begin{bmatrix} 0 & -1 \\ 1 & 0 \end{bmatrix}.$$

The intention behind these matrices is as follows: $\mathbf{Q}_3$ may do a transition labelled $\varepsilon, \varepsilon$ (*idle transition*); $\mathbf{Q}_3$ may receive one of four signal $00, 01, 10, 11$ and perform the given unitary transformations.

## 3.2 Alice and Bob

We now describe two classical $\mathbb{C}$-automata **Alice** and **Bob** which represent, respectively, the sender and the receiver of teletransport.

### 3.2.1 Alice

Let $X = \{x_1, x_2, x_3, x_{00}, x_{01}, x_{10}, x_{11}\}$. Then **Alice** is the classical $\mathbb{C}$-automaton with state space $\mathbb{C}^X$ with left interface $A_{Alice} = \{\varepsilon\}$ and right interface

$$B_{Alice} = \{\varepsilon, 00, 01, 10, 11\} \times \{\varepsilon, c, h, m_0, m_1\} \times \{\varepsilon, m_0, m_1\}$$
$$= B_{Alice,1} \times A_1 \times A_{22}.$$

and transformations as indicated in the diagram

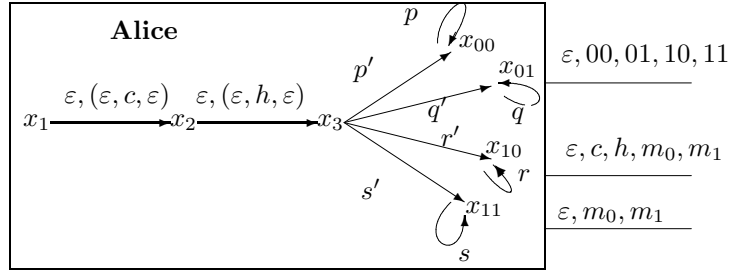

where $p' = \varepsilon, (\varepsilon, m_0, m_0)$, $q' = \varepsilon, (\varepsilon, m_0, m_1)$, $r' = \varepsilon, (\varepsilon, m_1, m_0)$, $s' = \varepsilon, (\varepsilon, m_1, m_1)$, $p = \varepsilon, (00, \varepsilon, \varepsilon)$, $q = \varepsilon, (01, \varepsilon, \varepsilon)$, $r = \varepsilon, (10, \varepsilon, \varepsilon)$, $s = \varepsilon, (11, \varepsilon, \varepsilon)$.

### 3.2.2 Bob

Let $Y = \{y_1, y_2\}$. Then **Bob** is the classical $\mathbb{C}$-automaton with state space $\mathbb{C}^Y$ with left interface $A_{Bob} = \{\varepsilon, 00, 01, 10, 11\} \times \{\varepsilon, 00, 01, 10, 11\}$ and right interface $B_{Bob} = \{\varepsilon\}$ and transformations relative to the standard basis $e_{y_1}, e_{y_2}$ having the following non-zero elements:

$$\varphi_{(\varepsilon,\varepsilon),\varepsilon}(e_{y_1}) = e_{y_1},$$
$$\varphi_{(00,00),\varepsilon}(e_{y_1}) = e_{y_2}, \varphi_{(01,01),\varepsilon}(e_{y_1}) = e_{y_2},$$
$$\varphi_{(10,10),\varepsilon}(e_{y_1}) = e_{y_2}, \varphi_{(11,11),\varepsilon}(e_{y_1}) = e_{y_2}.$$

## 4 The algebra of $\mathbb{C}$-automata

Now we define operations on $\mathbb{C}$-automata analogous (in a precise sense) to those defined in [9].

**Definition 4.1** *Given a $\mathbb{C}$-automata $\mathbf{Q}$ with left and right interfaces $A$ and $B$, state space $V$, and family of transformations $\varphi$, and $\mathbf{S}$ with interfaces $C$ and $D$, state space $W$, transformations $\psi$, the* parallel composite $\mathbf{Q} \otimes \mathbf{R}$ *is the $\mathbb{C}$-automaton which has state space $V \otimes W$, left interfaces $A \times C$, right interface $B \times D$, and transformations*

$$(\varphi \otimes \psi)_{(a,c),(b,d)} = \varphi_{a,b} \otimes \psi_{c,d}.$$

**Definition 4.2** *Given $\mathbb{C}$-automata $\mathbf{Q}$ with left and right interfaces $A$ and $B$, state space $V$, and family of transformations $\varphi$, and $\mathbf{R}$ with interfaces $B$ and $C$, state space $W$, and family of transformations $\psi$ the* series *(communicating* parallel) composite of $\mathbb{C}$-automata $\mathbf{Q} \circ \mathbf{R}$ has state space $V \otimes W$, left interfaces $A$, right interface $C$, and transition maps

$$(\varphi \circ \psi)_{a,c} = \sum_{b \in B} \phi_{a,b} \otimes \psi_{b,c}.$$

**Definition 4.3** *Given a relation $\rho \subset A \times B$ we define a $\mathbb{C}$-automaton $\rho$ as follows: it has state space $\mathbb{C}$. The transition matrices $\rho_{a,b}$ are $1 \times 1$ matrices, that is, complex numbers. Then $\rho_{a,b} = 1$ if $\rho$ relates $a$ and $b$, and $\rho_{a,b} = 0$ otherwise.*

Some special cases, all described in [9], have particular importance:

(i) the automaton corresponding to the identity function $1_A$, considered as a relation on $A \times A$ is called $1_A$;

(ii) the automaton corresponding to the diagonal function $\Delta : A \to A \times A$ (considered as a relation) is called $\Delta_A$; the automaton corresponding to the opposite relation of $\Delta$ is called $\nabla_A$.

(iii) the automaton corresponding to the function $twist : A \times B \to B \times A$ is called $twist_{A,B}$.

(iv) the automaton corresponding to the relation $\eta = \{(*, (a,a)); a \in A\} \subset \{*\} \times (A \times A)$ is called $\eta_A$; the automaton corresponding to the opposite of $\eta$ is called $\epsilon_A$.

## 4.1 The teleportation protocol

### 4.1.1 The protocol TP

Now the model of the teleportation protocol we consider is an expression in the algebra, involving also the automata $\mathbf{Q}_1, \mathbf{Q}_2, \mathbf{Q}_3$, **Alice**, and **Bob**. The protocol is

$$\mathbf{TP} = \mathbf{Alice} \circ (1_{A_3} \otimes ((\mathbf{Q}_1 \otimes 1_{A_{22}}) \circ \mathbf{Q}_2)) \circ (1_{A_3} \otimes \mathbf{Q}_3) \circ \mathbf{Bob}.$$

Notice that $(\mathbf{Q}_1 \otimes 1_{A_{22}}) \circ \mathbf{Q}_2$ and $\mathbf{Q}_3$ are quantum $\mathbb{C}$-automata.

As explained in [9], we may represent this system by the following diagram:

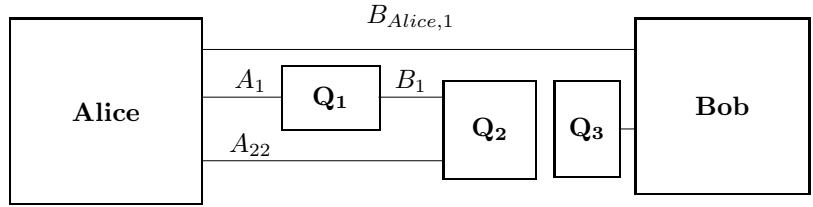

### 4.1.2 The behaviour of TP

Consider the following initial state of **TP**

$$x_1 \otimes (\alpha 0 + \beta 1) \otimes \frac{1}{\sqrt{2}}(0 \otimes 0 + 1 \otimes 1) \otimes y_1;$$

that is that state of $\mathbf{Q}_1$ is arbitrary and $\mathbf{Q}_2$ and $\mathbf{Q}_3$ are in Bell state. Since the combined system **TP** is closed it consists of a single linear transformation $\theta$ acting on the state space $\mathbb{C}^X \otimes \mathbb{C}^2 \otimes \mathbb{C}^2 \otimes \mathbb{C}^2 \otimes \mathbb{C}^Y$. A behaviour consists of a sequence of applications of $\theta$ to the initial state. However, in view of the construction of $\theta$ from parts, we may give a more explicit description of behaviours beginning in this initial state. In the following calculation it is critical that $\mathbb{C}^X$ and $C^Y$ break up into a direct sums $\mathbb{C} \oplus \mathbb{C} \oplus \cdots \oplus \mathbb{C}$ so that, using the distributive law of tensor over direct sum, **Alice** and **Bob** can do different actions on the qubits in different summands. This is entirely analogous to the use of sums and the distributive law in sequential programming, in particular in defining "if then else" [6],[20].

Simplifying the notation, writing for example 00 instead of $0 \otimes 0$, a four step behaviour is:

$$x_1 \otimes (\alpha 0 + \beta 1) \otimes \frac{1}{\sqrt{2}}(00 + 11) \otimes y_1$$

$$\mapsto x_2 \otimes \frac{1}{\sqrt{2}}(\alpha 000 + \alpha 011 + \beta 110 + \beta 101) \otimes y_1$$

$$\mapsto \frac{1}{2} x_3 \otimes (\alpha(0+1)00 + \alpha(0+1)11 + \beta(0-1)10 + \beta(0-1)01) \otimes y_1$$

$$= \frac{1}{2} x_3 \otimes (\alpha(000 + 100 + 011 + 111) + \beta(010 - 110 + 001 - 101)) \otimes y_1$$

$$\mapsto \frac{1}{2}(x_{00} \otimes (\alpha 000 + \beta 001) \otimes y_1 + x_{01} \otimes (\alpha 011 + \beta 010) \otimes y_1 +$$

$$x_{10} \otimes (\alpha 100 - \beta 101) \otimes y_1 + x_{11} \otimes (\alpha 111 - \beta 110) \otimes y_1)$$

$$\mapsto \frac{1}{2}(x_{00} \otimes (\alpha 000 + \beta 001) \otimes y_2 + x_{01} \otimes (\alpha 010 + \beta 011) \otimes y_2 +$$

$$x_{10} \otimes (\alpha 100 + \beta 101) \otimes y_2 + x_{11} \otimes (\alpha 110 + \beta 111) \otimes y_2)$$

$$= \frac{1}{2}(x_{00} \otimes 00 + x_{01} \otimes 01 + x_{10} \otimes 10 + x_{11} \otimes 11) \otimes (\alpha 0 + \beta 1) \otimes y_2.$$

### 4.2 The algebra of automata: equations

There is clearly much more to develop about the algebraic structure. We mention only that the constants $\Delta_A$, $\nabla_A$ satisfy the Frobenius equations [3], namely that

$$(\Delta_A \otimes 1_A) \circ (1_A \otimes \nabla_A) = \nabla_A \circ \Delta_A.$$

Notice that relations on $X$ also exist as closed classical automata with state space $\mathbb{C}^X$ and there the Frobenius equations are also satisfied, which fact has been used in axiomatizing classical data in [5]
.

# References

[1] S. Abramsky and B. Coecke, *A Categorical Semantics of Quantum Protocols*, in Proceedings of the 19th Annual IEEE Symposium on Logic in Computer Science: LICS 2004, IEEE Computer Society, 415–425, 2004.

[2] C. H. Bennett, G. Brassard, C. Crépeau, R. Jozsa, A. Peres, W. K. Wootters, *Teleporting an Unknown Quantum State via Dual Classical and Einstein-Podolsky-Rosen Channels*, Phys. Rev. Lett. 70, 1895-1899, 1993.

[3] A. Carboni, R.F.C. Walters, *Cartesian bicategories I*, Journal of Pure and Applied Algebra, 49, 11–32, 1987.

[4] Bob Coecke, Dusko Pavlovic, *Quantum measurements without sums*, archiv:quant-ph/0608072.v2, 2006.

[5] Bob Coecke, Eric O. Paquette, Dusko Pavlovic, *Classical and Quantum Structures*, Oxford University Computing Laboratory Report PRG-RR-08-02, 2008.

[6] C.C. Elgot, *Monadic computation and iterative algebraic theories*, Logic Colloquium 1973, Studies in Logic 80, North Holland, 175-230, 1975.

[7] L. de Francesco Albasini, N. Sabadini, R.F.C. Walters: *Cospan Span(Graphs): a compositional model for reconfigurable automata nets,* Developments and New Tracks in Trace Theory, Cremona, Italy, 9-11 October 2008.

[8] L. de Francesco Albasini, N. Sabadini, R.F.C. Walters: *The compositional construction of Markov processes*, arXiv:0901.2434, 2009.

[9] P. Katis, N. Sabadini, R.F.C. Walters, *Span(Graph): A categorical algebra of transition systems*, Proc. AMAST '97, SLNCS 1349, pp 307–321, Springer Verlag, 1997.

[10] P. Katis, N. Sabadini, R.F.C. Walters, *A formalisation of the IWIM Model*, in: Proc. COORDINATION 2000, LNCS 1906, 267–283, Springer Verlag, 2000.

[11] P. Katis, N. Sabadini, R.F.C. Walters, *Feedback, trace and fixed-point semantics*, Theoret. Informatics Appl. 36, pp 181–194, 2002.

[12] G.M. Kelly, M.L. Laplaza, *Coherence for compact closed categories*, J. Pure Appl. Algebra 19, 193-213, 1980.

[13] J. Kock, *Frobenius algebras and 2D topological Quantum Field Theories*, Cambridge University Press, 2004.

[14] A. Kondacs and J. Watrous, *On the power of quantum finite state automata*, Proceedings of the 38th Annual Symposium on Foundations of Computer Science, IEEE Computer Society, Los Alamitos, 66-75, 1997.

[15] José Meseguer, Ugo Montanari: *Petri Nets Are Monoids*, Information and Computation, Volume 88, 105–155, 1990.

[16] C. Moore, J. Crutchfield, *Quantum automata and quantum grammars*, Theoretical Computer Science, 237, 275–306, 2000.

[17] R. Penrose, *Applications of negative dimensional tensors*, In Combinatorial Mathematics and its Applications, Academic Press, 1971.

[18] Nicoletta Sabadini, Sebastiano Vigna, Robert F. C. Walters, *A Note on Recursive Functions*, Mathematical Structures in Computer Science 6(2): 127-139, 1996.

[19] Sebastiano Vigna, *On the Relations between Distributive Computability and the BS,* Model. Theor. Comput. Sci. 162(1): 5-21, 1996.

[20] R.F.C. Walters, *Categories and Computer Science*, Cambridge University Press, 1992.